\documentclass{ws-ijqi}
\usepackage{bbm,amsmath}
\begin{document}
%%%%%%%%%%%%%%%%%%
\def\set#1{{\cal #1}}\def\sH{\set{H}}
\def\Tr{\hbox{Tr}}
\def\sT{{\scriptscriptstyle T}}
\def\kket#1{|#1\rangle\rangle}
\def\bbra#1{\langle\langle #1|}
\newcommand{\ket}[1]{ |#1  \rangle}
\newcommand{\bra}[1]{ \langle #1|}
\newcommand{\dket}[1]{ | #1  \rangle\!\rangle}
%%%%%%%%%%%%%%%%%%
\markboth{Matteo G A Paris}{Local Quantum Estimation}
\title{QUANTUM ESTIMATION for QUANTUM TECHNOLOGY}
\author{MATTEO G A PARIS}
\address{Dipartimento di Fisica dell'Universit\`a di Milano, I-20133
Milano, Italia \\ CNSIM, Udr Milano, I-20133 Milano, Italia \\ ISI 
Foundation, I-10133 Torino, Italia}
\date{\today}
\maketitle
%%%%%%%%%%%%%%%%%%
\begin{abstract}
Several quantities of interest in quantum information, including
entanglement and purity, are nonlinear functions of the density matrix
and cannot, even in principle, correspond to proper quantum observables.
Any method aimed to determine the value of these quantities should
resort to indirect measurements and thus corresponds to a parameter
estimation problem whose solution, {\em i.e} the determination of the
most precise estimator, unavoidably involves an optimization procedure.
We review local quantum estimation theory and present explicit formulas
for the symmetric logarithmic derivative and the quantum Fisher
information of relevant families of quantum states.  Estimability of a
parameter is defined in terms of the quantum signal-to-noise ratio and
the number of measurements needed to achieve a given relative error. 
The connections between the optmization procedure and the geometry 
of quantum statistical models are discussed. Our analysis allows to 
quantify quantum noise in the measurements of non observable quantities
and provides a tools for the characterization of signals and 
devices in quantum technology.
\end{abstract}
%%%%%%%%%%%%%%%%%%
\section{Introduction}\label{s:intro}
Many quantities of interest in physics are not directly accessible,
either in principle or due to experimental impediments. This is
particolarly true for quantum mechanical systems where relevant
quantities like entanglement and purity are nonlinear functions of the
density matrix and cannot, even in principle, correspond to proper
quantum observables.
In these situations one should resort to indirect
measurements, inferring the value of the quantity of interest by
inspecting a set of data coming from the measurement of a different
obeservable, or a set of observables. This is basically a parameter
estimation problem which  may be properly addressed in the framework of
quantum estimation theory (QET) \cite{QET}, which provides
analytical tools to find the optimal measurement according to some
given criterion.  In turn, there are two main paradigms in QET:
{\em Global} QET looks for the POVM minimizing a suitable cost
functional, averaged over all possible values of the parameter to
be estimated.  The result of a global optimization is thus a single
POVM, independent on the value of the parameter. On the other
hand, {\em local} QET looks for the POVM maximizing the Fisher
information, thus minimizing the variance of the estimator, at a
fixed value of the parameter \cite{hel67,yl73,hk74,BC94,BC96}.  Roughly speaking,
one may expect local QET to provide better performances since the
optimization concerns a specific value of the parameter, with some
adaptive or feedback mechanism assuring the achievability of the
ultimate bound \cite{gil00}.  Global QET has been mostly applied
to find optimal measurements and to evaluate lower bounds on
precision for the estimation of parameters imposed by unitary
transformations. For bosonic systems these include single-mode
phase \cite{Hol79,Dar98}, displacement \cite{Hel74}, squeezing
\cite{Mil94,Chi06} as well as two-mode transformations, e.g.
bilinear coupling \cite{per01}. Local QET has been applied to the
estimation of quantum phase \cite{Mon06} and to estimation
problems with open quantum systems and non unitary processes
\cite{Sar06}: to finite dimensional systems \cite{Hot06}, to
optimally estimate the noise parameter of depolarizing
\cite{Fuj01} or amplitude-damping \cite{Zhe06}, and for continuous
variable systems to estimate the loss parameter of a quantum
channel \cite{Dau06,Gra87,Pol92,Mon07} as well as the position of a single 
photon \cite{fri07}. Recently, the geometric structure induced 
by the Fisher information itself has been exploited to give a 
quantitative operational interpretation for multipartite 
entanglement \cite{Boi08} and to assess quantum criticality as 
a resource for quantum estimation \cite{ZP07}. 
\par
In this paper we review local quantum estimation theory and present
explicit formulas for the symmetric logarithmic derivative and the
quantum Fisher information of relevant families of quantum states.  We
are interested in evaluating the ultimate bound on precision
(sensitivity), {\em i.e} the smallest value of the parameter that can be
discriminated, and to determine the optimal measurement achieving those
bounds.  Estimability of a parameter will be then defined in terms of
the quantum signal-to-noise ratio and the number of measurements needed
to achieve a given relative error.
\par
The paper is structured as follows. In the next Section we review local
quantum estimation theory and report the solution of the
optimization problem, {\em i.e.} the determination of the optimal 
quantum estimator in terms of the symmetric logarithmic derivative, 
as well as the ultimate bounds to precision in terms of the quantum
Fisher information. General formulas for the symmetric logarithmic 
derivative and the quantum Fisher information are derived. 
In Section \ref{s:qep} we address the quantification of estimability 
of a parameter put forward the quantum signal-to-noise ratio and
the number of measurements needed to achieve a given relative error
as the suitable figures of merit.
In Section \ref{s:exa} we present explicit formulas for sets of pure 
states and the generic unitary family. We also consider the multiparamer 
case and the problem of repametrization. In Section \ref{s:geo}
we discuss the connections between estimability of a set of parameters, 
the optmization procedure and the geometry of quantum statistical models.
Section \ref{s:out} closes the paper with some concluding remarks.
%%%%
\section{Local quantum estimation theory}\label{s:qet}
The  solution of a parameter estimation problem amounts to find
an estimator, {\em i.e} a mapping $\hat\lambda=\hat\lambda
(x_1,x_2,...)$ from the set $\chi$ of measurement outcomes into
the space of parameters. 
Optimal estimators in classical estimation theory are those 
saturating the Cramer-Rao inequality  \cite{Cra46} 
\begin{align}
{\mathrm{V}}(\lambda) \geq \frac{1}{M F(\lambda)} 
\label{eq:CramerRao}
\end{align}
which establishes a lower bound on the mean square error $V(\lambda)
= E_{\lambda} [(\hat\lambda(\{x\}) - \lambda)^2]$ of any estimator
of the parameter $\lambda$.
In Eq. (\ref{eq:CramerRao}) $M$ is the number of measurments 
and $F(\lambda)$ is the so-called 
Fisher Information (FI)
\begin{align}
F(\lambda) = \int\!\!dx\, p(x\vert \lambda) \left(
\frac{\partial \ln p(x\vert \lambda)}{\partial \lambda}
\right)^2 
= \int\!\!dx\, \frac{1}{p(x\vert \lambda)} \left(
\frac{\partial p(x\vert \lambda)}{\partial \lambda}
\right)^2 . \label{eq:ClassicalFisher}
\end{align}
where $p(x\vert\lambda)$ denotes the conditional probability of
obtaining the value $x$ when the parameter has the value $\lambda$.
For unbiased estimators,
as those we will deal with, the mean square error is equal to the
variance $\hbox{Var}(\lambda)= E_{\lambda} [\hat\lambda^2]  -
E_{\lambda} [\hat\lambda]^2$.
\par
When quantum systems are involved any estimation problem may be stated
by considering a family of quantum states $\varrho_\lambda$ which are
defined on a given Hilbert space ${\cal H}$ and labeled by  a parameter
$\lambda$ living on a $d$-dimensional manifold  ${\cal M}$, with the
mapping $\lambda \mapsto \varrho_\lambda$ providing a coordinate system.
This is sometimes referred to as a quantum statistical model.  The
parameter $\lambda$ does not, in general, correspond to a quantum
observable and our aim is to estimate its values through the measurement
of some observable on $\varrho_\lambda$.  In turn, a quantum estimator
$O_\lambda$ for $\lambda$ is a selfadjoint operator, which describe a
quantum measurement followed by any classical data processing performed
on the outcomes.  The indirect procedure of parameter estimation implies
an additional uncertainty for the measured value, that cannot be avoided
even in optimal conditions. The aim of quantum estimation theory is to
optimize the inference procedure by minimizing this additional
uncertainty.
\par
In quantum mechanics, according to the Born rule we have
$p(x\vert \lambda) = \Tr[\Pi_x \varrho_\lambda]$ where
$\left\{\Pi_x\right\}$, $\int\!\!dx\, \Pi_x = {\mathbbm I}$, 
are the elements of a positive operator-valued measure (POVM) and 
$\varrho_\lambda$ is the density operator parametrized by the quantity 
we want to estimate. Introducing the Symmetric Logarithmic
Derivative (SLD) $L_\lambda$ as the selfadjoint operator satistying 
the equation \begin{align}
\frac{L_\lambda \varrho_\lambda + \varrho_\lambda L_\lambda}{2} =
\frac{\partial \varrho_\lambda}{\partial \lambda} \label{eq:SLD}
\end{align}
we have that
$\partial_\lambda p(x\vert\lambda) = \Tr[ \partial_\lambda\varrho_\lambda \Pi_x]
= \hbox{Re}( \Tr[\varrho_\lambda \Pi_x L_\lambda ] )$. 
The Fisher Information (\ref{eq:ClassicalFisher}) is then rewritten as
\begin{align}
F(\lambda) = \int\!\!dx\, \frac{\hbox{Re}\left(\Tr\left[\varrho_\lambda \Pi_x
L_\lambda\right]\right)^2} {\Tr[\varrho_\lambda \Pi_x]}
\:. \label{eq:CQFisher}
\end{align}
For a given quantum measurement, {\em i.e.} a POVM $\{\Pi_x\}$, Eqs.
(\ref{eq:ClassicalFisher}) and (\ref{eq:CQFisher}) establish the 
classical bound on precision, which may be achieved by a proper
data processing, {\em e.g.} by maximum likelihood, which is
known to provide an asymptotically efficient estimator.
On the other hand, in order to evaluate the ultimate bounds to 
precision we have now to maximize the Fisher information over 
the quantum measurements. 
Following Refs. \cite{yl73,hk74,BC94,BC96} we have 
\begin{align}
F(\lambda)  & \leq  
\int\!\!dx\, \left|\frac{\Tr\left[\varrho_\lambda \Pi_x
L_\lambda\right]}{\sqrt{\Tr[\varrho_\lambda \Pi_x]}}\right|^2
\label{uno} \\ 
& = \int\!\!dx\, \left|
\Tr\left[ \frac{\sqrt{\varrho_\lambda}
\sqrt{\Pi_x}}{\sqrt{\Tr\left[\varrho_\lambda \Pi_x\right]}}\, \sqrt{\Pi_x} L_\lambda
\sqrt{\varrho_\lambda} \right] \right|^2 \nonumber \\
& \leq \int\!\!dx\, \Tr\left[\Pi_x L_\lambda \varrho_\lambda L_\lambda\right]
\label{due}\\
& =\Tr[L_\lambda \varrho_\lambda L_\lambda ] = \Tr[\varrho_\lambda L_\lambda^2]
\nonumber 
\end{align}
The above chain of inequalities prove that the Fisher information 
$F(\lambda)$ of any quantum measurement is bounded by the so-called 
\emph{Quantum Fisher Information} (QFI)
\begin{align}
F(\lambda) \leq H(\lambda) 
\equiv \Tr[\varrho_\lambda L_\lambda^2]
= \Tr[\partial_\lambda \varrho_\lambda L_\lambda]
\label{eq:QuantumFisher} 
\end{align}
leading the quantum Cramer-Rao bound 
\begin{align}
{\mathrm{Var}}(\lambda) \geq \frac{1}{M H(\lambda)} \label{QCR}
\end{align}
to the variance of any estimator. The quantum version of the 
Cramer-Rao theorem provides an ultimate bound: it does depend on 
the geometrical structure of the quantum statistical model and 
does not depend on the measurement. 
Optimal quantum measurements for the estimation of $\lambda$ 
thus corresponds to POVM with a Fisher information equal to the quantum Fisher
information, {\em i.e} those saturating both inequalities (\ref{uno}) 
and (\ref{due}). The first one is saturated when 
$\Tr[\varrho_\lambda \Pi_x L_\lambda]$ is a real number $\forall \lambda$. 
On the  other hand, Ineq. (\ref{due}) is based on the Schwartz inequality 
$|\Tr[A^\dag B]|^2 \leq \Tr[A^\dag A] \Tr[B^\dag B]$ applied to  
$A^\dag= \sqrt{\varrho_\lambda}
\sqrt{\Pi_x}/\sqrt{\Tr\left[\varrho_\lambda \Pi_x\right]}$ and 
$B=\sqrt{\Pi_x} L_\lambda \sqrt{\varrho_\lambda}$ and it is saturated when
\begin{equation}
\frac{\sqrt{\Pi_x}\sqrt{\varrho_\lambda}}
{\Tr\left[\varrho_\lambda \Pi_x\right]}
= 
\frac{\sqrt{\Pi_x} L_\lambda \sqrt{\varrho_\lambda}}
{\Tr[\varrho_\lambda \Pi_x L_\lambda]}\qquad \forall \lambda\:,
\label{opt}
\end{equation}
The operatorial condition in Eq. (\ref{opt}) is satisfied iff $\{\Pi_x\}$
is made by the set of projectors over the eigenstates of $L_\lambda$,
which, in turn, represents the optimal POVM to estimate the parameter
$\lambda$.  Notice, however, that $L_\lambda$ itself may not represent
the optimal observable to be measured. In fact, Eq. (\ref{opt})
determines the POVM and not the estimator {\em i.e} the function of the
eigenvalues  of $L_\lambda$. As we have already mentioned above, this 
corresponds to a classical
post-processing of data aimed to saturate the Cramer-Rao inequality
(\ref{eq:CramerRao}) and may be pursued by maximum likelihood, which is
known to provide an asymptotically efficient estimator. Using the fact
that $\Tr[\varrho_\lambda L_\lambda]=0$ an explicit form for the 
optimal quantum estimator is given by 
\begin{equation}
O_\lambda = \lambda {\mathbbm I} 
+ \frac{L_\lambda}{H(\lambda)}
\label{QEst}
\end{equation}
for which we have 
$$\Tr[\varrho_\lambda O_\lambda] = \lambda\,, \:\: \Tr[\varrho_\lambda
O_\lambda^2] = \lambda ^2 + \frac{\Tr[\varrho_\lambda
L^2_\lambda]}{H^2(\lambda)}\,,\:\: \hbox{and thus }
\langle \Delta O^2_\lambda \rangle = 1/H(\lambda)\:.$$
\par
Eq. (\ref{eq:SLD}) is Lyapunov matrix equation to be solved 
for the SLD $L_\lambda$. The general solution may be written as 
\begin{equation}
L_\lambda = 2 \! \int_0^\infty\!\! dt\, \exp\{-\varrho_\lambda t\}\, \partial_\lambda
\varrho_\lambda \exp\{-\varrho_\lambda t\}
\label{LE}
\end{equation}
which, upon writing $\varrho_\lambda$ in its eigenbasis
$\varrho_\lambda = \sum_n \varrho_n |\psi_n\rangle\langle\psi_n |$, 
leads to
\begin{equation}
L_\lambda = 2 \sum_{nm} \frac{\langle \psi_m| \partial_\lambda
\varrho_\lambda | \psi_n\rangle}{\varrho_n+ \varrho_m}
|\psi_m\rangle\langle\psi_n | \:, 
\label{LL}
\end{equation}
where the sums include only terms with $\varrho_n+\varrho_m \neq 0$.
The quantum Fisher information is thus given by
\begin{equation}
H (\lambda) = 2 \sum_{nm} \frac{\left|\langle \psi_m| \partial_\lambda
\varrho_\lambda | \psi_n\rangle\right|^2}{\varrho_n+ \varrho_m}\:,
\label{HH}
\end{equation}
or, in a basis independent form, 
\begin{equation}
H (\lambda) = 2 \! \int_0^\infty\!\! dt\, 
\Tr\left[
\partial_\lambda \varrho_\lambda 
\exp\{-\varrho_\lambda t\}\, 
\partial_\lambda \varrho_\lambda 
\exp\{-\varrho_\lambda t\}
\right]
\label{HHbi}\:.
\end{equation}
Notice that the SLD is defined only on the support of 
$\varrho_\lambda$ and that both the eigenvalues $\varrho_n$ and the
eigenvectors $|\psi_n\rangle$ may depend on the parameter.  
In order to separate the two contribution to the QFI we explicitly
evaluate $\partial_\lambda\varrho_\lambda$
\begin{equation}
\partial_\lambda\varrho_\lambda = \sum_p 
\partial_\lambda \varrho_p |\psi_p\rangle\langle \psi_p | +
\varrho_p |\partial_\lambda \psi_p\rangle\langle \psi_p | +
\varrho_p |\psi_p\rangle\langle\partial_\lambda  \psi_p | \label{der}
\end{equation}
The symbol $|\partial_\lambda \psi_n\rangle$ denotes 
the ket $|\partial_\lambda \psi_n\rangle = \sum_k 
\partial_\lambda \psi_{nk} | k\rangle$, where $\psi_{nk}$ 
are obtained expanding $|\psi_n\rangle$ in arbitrary basis 
$\{|k\rangle\}$  independent on $\lambda$. 
Since $\langle \psi_n | \psi_m\rangle = \delta_{nm}$ we
have $\partial_\lambda \langle \psi_n | \psi_m\rangle 
\equiv  \langle\partial_\lambda  \psi_n |\psi_m\rangle +
\langle \psi_n |\partial_\lambda \psi_m\rangle = 0$
and therefore 
$$ \hbox{Re} \langle\partial_\lambda  \psi_n |\psi_m\rangle = 0 
\qquad \langle\partial_\lambda  \psi_n |\psi_m\rangle = - 
\langle \psi_n |\partial_\lambda \psi_m\rangle = 0\:. $$
Using Eq. (\ref{der}) and the above identities we have  
\begin{equation}
L_\lambda = 
\sum_p \frac{\partial_\lambda \varrho_p}{\varrho_p} 
|\psi_p\rangle\langle \psi_p | +
2 \sum_{n\neq m} \frac{\varrho_n - \varrho_m}{\varrho_n + \varrho_m}
\langle \psi_m |\partial_\lambda \psi_n\rangle  
\: |\psi_m\rangle\langle \psi_n | 
\label{Le}
\end{equation}
and in turn 
\begin{equation}
\label{He}
H (\lambda) = 
\sum_p \frac{\left(\partial_\lambda \varrho_p\right)^2}{\varrho_p} 
+ 2 \sum_{n\neq m} \sigma_{nm}  
\left|\langle \psi_m |\partial_\lambda \psi_n\rangle  \right|^2
\end{equation}
where   
\begin{equation}\label{sg1} 
\sigma_{nm} = \frac{(\varrho_n - \varrho_m)^2}{\varrho_n+\varrho_m} + 
\hbox{any antisymmetric term}\:,
\end{equation}
as for example 
\begin{equation}\label{sg2} 
\sigma_{nm} = 2 \varrho_n \frac{\varrho_n - \varrho_m}{\varrho_n+\varrho_m} 
\qquad 
\sigma_{nm} = 2 \varrho_n \left(\frac{\varrho_n 
- \varrho_m}{\varrho_n+\varrho_m}\right)^2 
\end{equation}
The first term in Eq. (\ref{He}) represents the classical Fisher
information of the distribution $\{\varrho_p\}$ whereas the second 
term contains the truly quantum contribution. The second term vanishes
when the eigenvectors of $\varrho_\lambda$ do not depend. In this case
$[\varrho_\lambda, \partial_\lambda\varrho_\lambda]=0$ and Eq. (\ref{LE})
reduces to $L_\lambda = \partial_\lambda \log \varrho_\lambda$.
\par 
Finally, upon substituting the above Eqs. in Eq. (\ref{QEst}), 
we obtain the corresponding optimal quantum estimator 
\begin{align}
O_\lambda = 
\sum_p \left( \lambda + \frac{\partial_\lambda \varrho_p}{\varrho_p}
\right)
|\psi_p\rangle\langle \psi_p | +
\frac{2}{H(\lambda)} \sum_{n\neq m} \frac{\varrho_n - \varrho_m}{\varrho_n + \varrho_m}
\langle \psi_m |\partial_\lambda \psi_n\rangle  
\: |\psi_m\rangle\langle \psi_n |\:. 
\end{align}
%%%
So far we have considered the case of a parameter with a fixed given 
value. A question arises on whether a bound for estimator variance may 
be established also for a parameter having an {\em a priori} distribution 
$z(\lambda)$. The answer is positive and given by the Van Trees inequality 
\cite{Van67,Gil95} which provides a bound for the average variance 
$$
\overline{\hbox{Var}(\lambda)}
= \int\!\! dx\!\! \int\!\! d\lambda \:z(\lambda)\:
\left[\hat\lambda(\{x\})-\lambda)\right]^2
$$ of any unbiased estimator of the random parameter $\lambda$.
Van Trees inequality states that
\begin{equation}
\overline{\hbox{Var}(\lambda)} \geq \frac1{Z_F}
\end{equation}
where the generalized Fisher information $Z_F$ is given by
\begin{equation} 
Z_F= \int\!\! dx\!\! \int\!\! d\lambda \: p(x,\lambda) \left[\partial_\lambda
\log p(x,\lambda)\right]^2 \:, \label{ZZ}
\end{equation} 
$p(x,\lambda)$ being the joint probability distribution of the outcomes
and the parameter of interest.  Upon writing the joint distribution 
as $p(x,\lambda) = p(x|\lambda) z(\lambda)$ Eq. (\ref{ZZ}) may be
rewritten as 
\begin{equation} 
Z_F= \int\!\! d\lambda\: z(\lambda)\: F (\lambda)  + M
\int\!\! d\lambda\: z(\lambda) \left[\partial_\lambda 
\log z(\lambda)\right]^2 \:.
\label{ZZ1}
\end{equation} 
Eq. (\ref{ZZ1}) says that the generalized Fisher information is the sum
of two terms, the first is simply the average of the Fisher information
over the {\em a priori} distribution whereas the second term is the
Fisher information of the priori distribution itself.  As expecteed, in 
the asymptotic limit of many measurements the {\em a priori} distribution 
is no longer relevant.  The
quantity $Z_F$ is upper bounded by the analogue expression $Z_H$ where
the average of the Fisher information is replaced by the average of the
QFI $H(\lambda)$ The resulting quantum Van Trees bound may be easily
written as
\begin{equation}
\overline{\hbox{Var}(\lambda)} \geq \frac1{Z_H}\:.
\end{equation}
%%%%
\section{Estimability of a parameter}\label{s:qep}
A large signal is easily estimated whereas a quantity with 
a vanishing value may be inferred only if the corresponding
estimator is very {\em precise} {\em i.e} characterized by
a small variance. This intuitive statement indicates that 
in assessing the performances of an estimator and, in turn, 
the overall estimability of a parameter, the relevant figure 
of merit is the scaling of the variance with the mean value
rather than its absolute value. This feature may be quantified 
by means of the signal-to-noise ratio (for a single measurement) 
$$ 
R_\lambda = \frac{\lambda^2}{\hbox{Var}(\lambda)}
$$
which is larger for better estimators.
Using the quantum Cramer-Rao bound one easily 
derives that the signal-to-noise ratio of any estimator 
is bounded by the quantity
$$
R_\lambda \leq Q_\lambda \equiv \lambda^2 H(\lambda)
$$
which we refer to as the quantum signal-to-noise ratio.
We say that a given parameter $\lambda$ is effectively estimable 
quantum-mechanically when the corresponding $Q_\lambda$ is large. 
\par
Upon taking into account repeated measurements we have that 
the number of measurements leading to a $99.9 \%$ ($3\sigma$) 
confidence interval corresponds to a relative
error 
$$
\delta^2 = \frac{9 \hbox{Var}(\lambda)}{M\lambda^2} = \frac{9}{M}
\frac1Q_\lambda = \frac{9}{M\lambda^2 H(\lambda)}
$$
Therefore, the number of measurements needed to achieve a $99.9\%$
confidence interval with a relative errro $\delta$ scales as
$$
M_\delta = \frac9{\delta^2} \frac1Q_\lambda
$$
In other words, a vanishing $Q_\lambda$ implies a diverging number of
measurements to achieve a given relative error, whereas a finite value
allows estimation with arbitrary precision at finite number of
measurements. 
%%%
\section{Examples}\label{s:exa}
In this section we provide explicit evaluation of the symmetric
logarithmic derivative and the quantum Fisher information for relevant
families of quantum states, including sets of pure states and the
generic unitary family. We also consider the multiparamer case and
the problem of repametrization.
\subsection{Unitary families and the pure state model}
Let us consider the case where the parameter of interest is 
the amplitude of a unitary perturbation imposed to a given
initial state $\varrho_0$. The family of quantum states we are dealing
with may be expressed as $\varrho_\lambda = U_\lambda \varrho_0
U^\dag_\lambda$ where $U_\lambda = \exp\{-i\lambda G\}$ is a unitary
operator and $G$ is the corresponding Hermitian generator.
Upon expanding the unperturbed state in its eigenbasis $\varrho_0 = \sum
\varrho_n |\varphi_n\rangle\langle\varphi_n |$  we have $\varrho_\lambda
= \sum_n \varrho_n |\psi_n\rangle\langle\psi_n |$ where 
$|\psi_n\rangle = U_\lambda |\varphi_n\rangle$. As a consequence 
we have 
$$\partial_\lambda \varrho_\lambda = i U_\lambda 
[G,\varrho_0]U^\dag_\lambda\,.$$ 
and the SLD is may be written as 
$L_\lambda = U_\lambda L_0 U^\dag_\lambda$ where  $L_0$ is given by 
\begin{align}
L_0 & = 2 i \sum_{n,m} \frac{\langle\varphi_m| [G,\varrho_0] |
\varphi_n\rangle}{\varrho_n+\varrho_m}\,
|\varphi_n\rangle\langle\varphi_m | = 
2 i \sum_{n\neq m}\langle\varphi_m| G |
\varphi_n\rangle \frac{\varrho_n-\varrho_m}{\varrho_n+\varrho_m}\,
|\varphi_n\rangle\langle\varphi_m |\,.
\end{align}
The corresponding quantum Fisher information 
is independent on the value of parameter and may be written 
in compact form as 
$$
H 
= \hbox{Tr}\left[ \varrho_0\, L_0^2 \right]
= \hbox{Tr}\left[ \varrho_0\, [L_0,G] \right]
= \hbox{Tr}\left[ L_0\, [G,\varrho_0] \right]
= \hbox{Tr}\left[ G\, [\varrho_0, L_0] \right]
$$
or, more explicitly, as 
$$
H = 2 \sum_{n\neq m} \sigma_{nm} G^2_{nm}
$$
where the elements $\sigma_{nm}$ are given in Eq. (\ref{sg1}), or 
equivalently (\ref{sg2}), and $G_{nm}=\langle\varphi_n | G |
\varphi_m\rangle=\langle\psi_n | G |
\psi_m\rangle$ denote the matrix element
of the generator $G$ in either the eigenbasis of $\varrho_0$ 
or $\varrho_\lambda$.
\par
For a generic family of pure states we have $\varrho_\lambda = 
|\psi_\lambda\rangle\langle\psi_\lambda |$. Since $\varrho_\lambda^2 =
\varrho_\lambda$ we have $\partial_\lambda\varrho_\lambda
= \partial_\lambda \varrho_\lambda\,\varrho_\lambda
+ \varrho_\lambda \partial_\lambda \varrho_\lambda$ and thus
$L_\lambda = 2 \partial_\lambda \varrho_\lambda
=|\psi_\lambda\rangle\langle\partial_\lambda\psi_\lambda |
+|\partial_\lambda\psi_\lambda\rangle\langle\psi_\lambda |
$.
Finally we have 
\begin{equation}\label{pure} 
H(\lambda) = 4 
\left[
\langle\partial_\lambda\psi_\lambda
|\partial_\lambda\psi_\lambda\rangle
+ \left( \langle\partial_\lambda\psi_\lambda |
\psi_\lambda\rangle
\right)^2
\right]
\end{equation}
\par
For a unitary family of pure states $|\psi_\lambda\rangle = U_\lambda
|\psi_0\rangle$ we have 
\begin{align}
|\partial_\lambda \psi_\lambda\rangle & = - i G 
U_\lambda |\psi_0\rangle = -i G |\psi_\lambda\rangle
\:,\nonumber \\ 
\langle\partial_\lambda\psi_\lambda
|\partial_\lambda\psi_\lambda\rangle & = 
\langle\psi_0| G^2 | \psi_0\rangle 
\:, \nonumber \\  
\langle\partial_\lambda\psi_\lambda |
\psi_\lambda\rangle & = -i 
\langle\psi_0| G | \psi_0\rangle 
\nonumber\:. 
\end{align}
The quantum Fisher information thus reduces to the simple form
\begin{equation}
H = 4  \langle\psi_0| \Delta G^2 | \psi_0\rangle 
\label{HUnPure}
\end{equation}
which is independent on $\lambda$ and proportional to the fluctuations
of the generator on the unperturbed state. Using Eq. (\ref{HUnPure})
the quantum Cramer-Rao bound in (\ref{QCR}) rewrites in the appealing
form \cite{mac06}
\begin{equation}
\hbox{Var}(\lambda) \langle\Delta G^2 \rangle \geq \frac1{4M}\,,
\label{parUN}
\end{equation}
which represents a parameter-based uncertainty relation which applies 
also  when the shift parameter $\lambda$ in the unitary $U_\lambda =
e^{- i \lambda G}$ {\em does not} correspond to the observable canonically
conjugate to $G$. 
When the unperturbed state is not pure the QFI may be written as
\begin{align}
H & = 4\,\Tr\left[\Delta G^2 \varrho_0\right]
+ 4 \sum_n \varrho_n \langle \varphi_n | \langle G\rangle ^2 - 2 
G K^{(n)} G | \varphi_n\rangle
\label{HUnMix} \\ 
K^{(n)} &= \sum_{m} \frac{\varrho_m}{\varrho_n+\varrho_m} 
|\varphi_m\rangle\langle\varphi_m | \stackrel{\varrho_0 \rightarrow 
|\varphi_0\rangle\langle\varphi_0 |}{\longrightarrow}
\frac12|\varphi_0\rangle\langle\varphi_0 |
\end{align}
and Eq. (\ref{parUN}) becomes 
\begin{align}
\hbox{Var}(\lambda) \langle\Delta G^2 \rangle \geq \frac1{4M}
\left[1+ \sum_n \varrho_n \langle \varphi_n | \langle G\rangle ^2 -2 
G K^{(n)} G | \varphi_n\rangle
\right]^{-1}\:. \label{parUNm}
\end{align}
The second term in Eqs. (\ref{HUnMix}) and (\ref{parUNm}) thus
represents the {\em classical} contribution to uncertainty due to
the mixing of the initial signal.
\par
As we have seen, for unitary families of quantum states the 
QFI is independent on the value of the parameter. As a consequence
the quantum signal-to-noise ratio
$Q_\lambda$ vanishes for vanishing $\lambda$ and thus the number of
measurements needed to achieve a relative error $\delta$ diverges as 
$M_\delta \sim (\delta \lambda)^{-2}$.
\subsection{Quantum operations}
Let us now consider a family of quantum states obtained from a given
inital state $\varrho_0$ by the action of a generic quantum operation
$\varrho_\lambda = {\cal E}_\lambda (\varrho_0) = \sum_k M_{k\lambda}
\varrho_0 M^\dag_{k\lambda}$. Upon writing the initial and the evolved 
states in terms of their eigenbasis 
$\varrho_0 = \sum_s \varrho_{0s} |\varphi_s\rangle\langle\varphi_s |$,
$\varrho_\lambda = \sum_s \varrho_{n} |\psi_n\rangle\langle\psi_n |$ 
we may evaluate the SLD and the
quantum Fisher information using Eqs. (\ref{LL}) and (\ref{HH}) where
\begin{align}
\varrho_n  =& \sum_{ks} \varrho_{0s} \left|
\langle\psi_n|M_{k\lambda}|\varphi_s\rangle\right|^2 \\
\langle\psi_m|\partial_\lambda\varrho_\lambda |\psi_n\rangle  =& \sum_{ks} \varrho_{0s}
\left[ 
\langle\psi_m|\partial_\lambda M_{k\lambda}|\varphi_s\rangle
\langle\varphi_s|M_{k\lambda}^\dag|\psi_n\rangle
\right. \nonumber \\ &+ \left. 
%+
\langle\psi_m|M_{k\lambda}|\varphi_s\rangle
\langle\varphi_s|\partial_\lambda M_{k\lambda}^\dag|\psi_n\rangle
\right] \:.
\end{align}
For a pure state at the input $\varrho_0=|\psi_0\rangle\langle\psi_0 |$
the above equation rewrites without the sum over $s$.
%%%
\subsection{Multiparametric models and reparametrization}
In situations where more than a parameter is involved the family 
of quantum states  $\varrho_{\boldsymbol\lambda}$ depends on a set
${\boldsymbol\lambda}=\{ \lambda_\mu \}$, $\mu=1,\dots,N$. 
In this cases the relevant object in the estimation problem 
is given by the so-called quantum Fisher information matrix, 
whose elements are defined as
\begin{align}
{\boldsymbol H}({\boldsymbol\lambda})_{\mu\nu} =& 
\Tr\left[\varrho_{{\boldsymbol\lambda}} \frac{L_\mu L_\nu +L_\nu
L_\mu}{2}\right] 
= \Tr[\partial_\nu \varrho_{\boldsymbol \lambda} L_\mu]
= \Tr[\partial_\mu \varrho_{\boldsymbol \lambda} L_\nu]
\nonumber \\= & 
\sum_n \frac{(\partial_\mu \varrho_n) (\partial_\nu \varrho_n)}{\varrho_n}
+ \sum_{n\neq m}
\frac{(\varrho_n-\varrho_m)^2}{\varrho_n + \varrho_m} \times \nonumber \\
&\times \left[\langle \psi_n|\partial_\mu \psi_m\rangle
\langle \partial_\nu \psi_m| \psi_n \rangle + 
\langle \psi_n|\partial_\nu \psi_m\rangle
\langle \partial_\mu \psi_m| \psi_n \rangle\right] 
\end{align}
where $L_\mu$ is the SLD corresponding to the parameter $\lambda_\mu$.
The Cramer-Rao theorem for multiparameter estimation says that the inverse 
of the Fisher matrix provides a lower bound  
on the covariance matrix $\hbox{Cov}[\boldsymbol\gamma]_{ij}= 
\langle \lambda_i \lambda_j \rangle -
\langle \lambda_i\rangle\langle\lambda_j\rangle$, {\em i.e}
$$
\hbox{Cov}[\boldsymbol\gamma] \geq \frac1M 
{\boldsymbol H}({\boldsymbol\lambda})^{-1}
$$
The above relation is a matric inequality and the corresponding
bound may not be achievable achievable in a multiparameter estimation. 
On the other hand, the diagonal elements of the inverse Fisher matrix provide
achievable bounds for the variances of single parameter estimators 
{\em at fixed value} of the others, in formula
\begin{align}
{\mathrm{Var}}(\lambda_\mu) = \gamma_{\mu\mu} \geq \frac1M ({\boldsymbol
H}^{-1})_{\mu\mu}. 
\label{eq:QCRMulti}
\end{align}
Of  course, for a diagonal Fisher matrix ${\mathrm{Var}}(\lambda_\mu) 
\geq 1/{\boldsymbol H}_{\mu\mu}$. 
\par 
Let us now suppose that the quantity of interest $g$ is a known function 
$g(\boldsymbol\lambda)$ of the parameters used to label the family of states. 
In this case we need to reparametrize the familiy with a new set of parameters 
$\widetilde{{\boldsymbol\lambda}}= \{\widetilde{\lambda}_j=
\widetilde{\lambda}_j({\boldsymbol\lambda})$ that includes the 
quantity of interest, {\em e.g} $\widetilde{\lambda}_1 \equiv
g(\boldsymbol\lambda)$.  Since  
$\widetilde{\partial}_\mu = \sum_\nu B_{\mu\nu} \partial_\nu$  
where
$B_{\mu\nu} = \partial\lambda_\nu/\partial \widetilde{\lambda}_\mu$
it is easy to prove that
$$\widetilde{L}_\mu = \sum_\nu B_{\mu\nu}L_\nu 
\qquad\widetilde{\boldsymbol H} = {\boldsymbol B} {\boldsymbol H}
{\boldsymbol B}^T \:.$$
The ultimate precision on the estimation of $g$ at fixed values
of the other parameters is thus given by 
$$
\hbox{Var}(g) \geq \frac{1}{M} (\widetilde{\boldsymbol H}^{-1})_{11}
$$
%%%
\section{Geometry of quantum estimation}\label{s:geo}
The estimability of a set of parameters labelling the family 
of quantum states $\{\varrho_{\boldsymbol\lambda}\}$ is naturally 
related to the distinguishability of the states within the quantum
statistical model {\em i.e.} with the notions of distance.
On the manifold of quantum states, however, different distances
may be defined and a question arises on which of them captures
the notion of estimation measure. As it can be easily proved 
it turns out that the Bures distance 
\cite{bur69,uhl76,jos94,hub92,sla96,hal98,dit99}
is the proper quantity to 
be taken into account.  This may be seen as follows.
The Bures distance between two density matrices is defined as
$D_B^2 (\varrho,\sigma)= 2[1-\sqrt{F(\varrho,\sigma)}]$ where 
$F(\varrho,\sigma) = \left(
\Tr\left[\sqrt{\sqrt{\varrho}\sigma\sqrt{\varrho}}\right]\right)^2$
is the fidelity. The Bures metric $g_{\mu\nu}$ is obtained upon 
considering the distance for two states obtained by an infinitesimal 
change in the value of the parameter
$$
d^2_{B} = D^2_B (\varrho_{\boldsymbol \lambda}, \varrho_{\boldsymbol
\lambda + d \boldsymbol\lambda})
= g_{\mu\nu} d\lambda_\mu d\lambda_\nu\:.
$$
By explicitly evaluating the Bures distance \cite{som03} one arrives at
$g_{\mu\nu}= \frac14 {\boldsymbol H}_{\mu\nu} (\boldsymbol\lambda)$, {\em i.e.}
the Bures metric is simply proportional to the QFI, which itself 
is symmetric, real and positive semidefinite, {\em i.e.} 
represents a metric for the manifold underlying the quantum
statistical model.
Indeed, a large QFI for a given $\lambda$ implies that the quantum 
states $\varrho_{\boldsymbol \lambda}$ and $\varrho_{\boldsymbol
\lambda + d \boldsymbol\lambda}$ should be statistically distinguishable
more effectively than the analogue states for a value $\lambda$
corresponding to smaller QFI. In other words, one confirms 
the intuitive picture in which optimal estimability (that
is, a diverging QFI) corresponds to quantum states that are sent
far apart upon infinitesimal variations of the parameters. 
\begin{figure}[h]
\centerline{\includegraphics[width=0.9\textwidth]{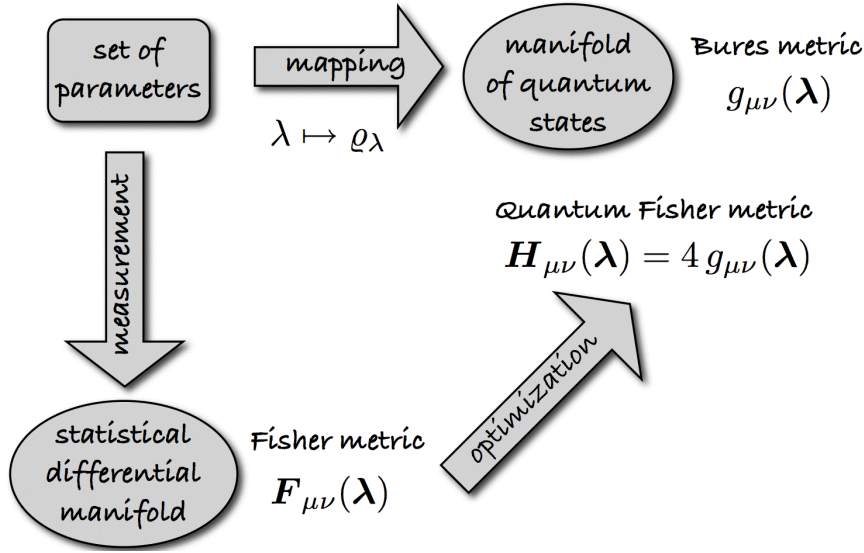}}
\caption{Geometry of quantum estimation}
\label{f:geo}
\end{figure} \\
The structures described above are pictorially described in Fig. 
\ref{f:geo}. The idea is that any measurement aimed to estimate 
the parameters ${\boldsymbol \lambda}$ 
turns the set of parameters into a statistical differential 
manifold endowed with the Fisher metric $\boldsymbol F_{\mu\nu}
({\boldsymbol\lambda})$.  On the other hand, when the parameters 
are mapped into the manifold of quantum states the statistical 
distance is expressed in terms of the Bures metric. The connection 
between the two constructions is provided by the optimization of 
the estimation procedure over quantum measurements, which 
shows that the Quantum Fisher metric $\boldsymbol H_{\mu\nu}
(\boldsymbol \lambda)$ is the bound to $\boldsymbol F_{\mu\nu}
({\boldsymbol\lambda})$ and coincides, apart from a factor four, 
with the Bures metric.
%%%
\section{Conclusions and outlooks}\label{s:out}
As a matter of fact, there are many quantities of interest that do not
correspond to any quantum observable. Among these, we mention the amount
of entanglement and the purity of a quantum state and the coupling
constant of an interaction Hamiltonian or a quantum operation. In these
situations, the values of the quantity of interest can be indirectly
inferred by an estimation procedure, {\em i.e.} by measuring one or more
proper observables, a quantum estimator, and then manipulating the
outcomes by a suitable classical processing. 
\par
In this paper, upon exploiting the geometric theory of quantum
estimation, we have described a general method to solve a quantum
statistical model, {\em i.e} to find the optimal quantum estimator and
to evaluate the corresponding bounds to precision.  To this aim we used
the quantum Cramer-Rao theorem and the explicit evaluation of the
quantum Fisher information matrix.  We have derived the explicit form of
the optimal observable in terms of the symmetric logarithmic derivative
and evaluated the corresponding bounds to precision, which represent the
ultimate bound posed by quantum mechanics to the precision of parameter
estimation. For unitary families of quantum states the bounds may
expressed in the form of a parameter-based uncertainty relation.  
\par
The analysis reported in this paper has a fundamental interest and 
represents a relevant tool in the design of realistic
quantum information protocols. The approach here outlined is currently
being applied to the estimation of entanglement \cite{ee08} and the
coupling constant of an interaction Hamiltonian \cite{ZP07,MK08}.
%%%
\section*{Acknowledgments}
The author thanks Paolo Giorda, Alex Monras, Paolo Zanardi, Marco 
Genoni, Michael Korbman, Carmen Invernizzi and Stefano Olivares for 
stimulating discussions.
%%%%%%%%%%%%%%%%%%

%%%%%%%%%%%%%%%%%%
\end{document}